\documentclass[apj,numberedappendix,appendixfloats]{emulateapj}

\usepackage{bm}
\usepackage{amsmath}
\usepackage{amssymb}
\usepackage{mathrsfs}

\input{colordvi.tex}

\newcommand{\simgt}{\lower.5ex\hbox{$\; \buildrel > \over \sim \;$}}
\newcommand{\simlt}{\lower.5ex\hbox{$\; \buildrel < \over \sim \;$}}

\shorttitle{Simulations of Wide-Field Weak Lensing Surveys II}
\shortauthors{Sato, Takada, Hamana and Matsubara}

\begin{document}

\title{Simulations of Wide-Field Weak Lensing Surveys II:
 Covariance Matrix of Real Space Correlation Functions }

\author{Masanori Sato\altaffilmark{1},
 Masahiro Takada\altaffilmark{2},
 Takashi Hamana\altaffilmark{3}
 and Takahiko Matsubara\altaffilmark{1,4}}

\affil{\altaffilmark{1} Department of Physics, 
 Nagoya University, Nagoya 464--8602, Japan}
\affil{\altaffilmark{2} Institute for the Physics and Mathematics of the
 Universe (IPMU), University of Tokyo, Chiba 277--8582, Japan}
\affil{\altaffilmark{3} National Astronomical Observatory of Japan,
  Tokyo 181--8588, Japan}
\affil{\altaffilmark{4} Kobayashi-Maskawa Institute for the Origin of
 Particles and the Universe, 
 Nagoya University, Nagoya 464--8602, Japan}
\email{masanori@a.phys.nagoya-u.ac.jp}

\begin{abstract}
 Using 1000 ray-tracing simulations for a
 $\Lambda$-dominated cold dark
 model in \cite{2009ApJ...701..945S}, we study the covariance matrix of
 cosmic shear correlation functions, which is the standard statistics
 used in the previous measurements. The shear correlation function of a
 particular separation angle is affected by Fourier modes over a wide
 range of multipoles, even beyond a survey area, which complicates the 
 analysis of the covariance matrix. To overcome such obstacles
 we first construct 
 Gaussian shear simulations from the 1000 realizations, and then use the
 Gaussian simulations to disentangle the Gaussian covariance
 contribution to the covariance matrix we measured from the original
 simulations.  We found that 
an analytical formula of Gaussian
 covariance overestimates the covariance amplitudes due to an
 effect of finite survey area. 
Furthermore, the clean separation of the
 Gaussian covariance allows to examine the non-Gaussian covariance
 contributions as a function of separation angles and source
 redshifts.  For upcoming surveys with typical source redshifts of
 $z_s=0.6$ and 1.0, the non-Gaussian contribution to the diagonal
 covariance components at 1 arcminute scales is greater than the
 Gaussian contribution by a factor of 20 and 10, respectively. 
 Predictions based on the halo model 
 qualitatively well reproduce the simulation
 results, however show a sizable disagreement in the 
 covariance amplitudes. By combining these simulation results we develop
 a fitting formula to the covariance matrix for a survey with
 arbitrary area coverage, taking into account effects of the finiteness
 of survey area on the Gaussian covariance.
\end{abstract}


\section{Introduction}
Weak gravitational lensing by intervening large scale structure, the
so-called cosmic shear,  
provides
a powerful
probe of dark matter and dark energy.  Since its first detections by
various
groups~\citep{2000MNRAS.318..625B,2000astro.ph..3338K,
2000A&A...358...30V,2000Natur.405..143W},
substantial progress has been made on both theoretical and observational
sides \citep[e.g.][]{2003ApJ...597...98H,2005MNRAS.359.1277M,2006ApJ...644...71J,2006A&A...452...51S,2008A&A...479....9F,2010A&A...516A..63S}.

Weak lensing has the highest potential to constrain 
properties of dark energy among other cosmological
observations, such as type Ia supernovae~\citep[e.g.][]{1998AJ....116.1009R,1999ApJ...517..565P,2009ApJ...700.1097H},
baryon acoustic
oscillations~\citep[e.g.][]{2005ApJ...633..560E,2008ApJ...676..889O},
galaxy clusters~\citep[e.g.][]{2009ApJ...692.1060V,2010MNRAS.406.1759M},
if the systematic errors are well under control~\citep{2006astro.ph..9591A,2009arXiv0901.0721A,2009PhRvD..80b3003J}.
The growth rate of mass clustering can be measured by ``lensing
tomography''~\citep[e.g.][]{1999ApJ...522L..21H,2002PhRvD..65f3001H,2004MNRAS.348..897T}
which in turn provides tight constraints on the equation of
state of dark energy.
For this purpose, a number of ambitious wide-field surveys have been
proposed, such as Subaru Hyper Suprime-Cam Survey~\citep{2006SPIE.6269E...9M},
the Panoramic Survey Telescope \& Rapid Response System~(Pan-STARRS\footnote{http://pan-starrs.ifa.hawaii.edu/public/}),
 the Dark Energy Survey~(DES\footnote{http://www.darkenergysurvey.org/}),
 the Large Synoptic Survey Telescope~(LSST\footnote{http://www.lsst.org/}),
 the Wide-Field Infrared Survey Telescope~(WFIRST\footnote{http://wfirst.gsfc.nasa.gov/}),
 and Euclid~\citep{2010arXiv1001.0061R}.

To attain the full potential of 
upcoming lensing surveys, 
it is essential to analyze data with appropriate statistical methods 
as well as to use sufficiently accurate theoretical models for the 
power spectrum and/or two-point correlation function and for 
the covariance matrix \citep[e.g.][for a recent development of lensing
power spectrum measurement method]{2011MNRAS.412...65H}.
Most of the useful cosmological information contained in the cosmic shear
signal lies in small scales that are affected by nonlinear clustering.
Therefore, non-Gaussian errors can be significant in weak lensing
measurements as indicated by several
studies~\citep{2000ApJ...537....1W,2001ApJ...554...56C,2007MNRAS.375L...6S,2009arXiv0905.0501D,2009A&A...502..721E,2009ApJ...701..945S,2009MNRAS.395.2065T,2010PhRvD..81l3015L,2010A&A...514A..79P,2011ApJ...729L..11S}
and also may cause biases in 
the best-fit parameters ~\citep{2009A&A...504..689H,2009PhRvD..79b3520I}.
Furthermore, future high-precision measurements may require
adequate statistical methods,
i.e. accurate likelihood function, of weak lensing 
in estimating cosmological parameters~\citep{2010PhRvL.105y1301S,2011PhRvD..83b3501S}.

In the first paper of a series of our works~\citep{2009ApJ...701..945S}, 
we studied the non-Gaussian effects on the covariance matrix of the 
cosmic shear power spectrum using 1000
realizations of ray-tracing simulations.
In this paper, we study the non-Gaussian effects on the covariance
matrix of the cosmic shear correlation function which is the most
conventionally used statistical 
method in the previous measurements.

This paper is organized as follows. In \S~\ref{sec:basics} we briefly
review the basics of the cosmic shear correlation function and its
covariance. In \S~\ref{sec:results} we show the main results, and
develop the fitting formula to compute the covariance matrix of cosmic
shear correlation functions for a survey of arbitrary area.
\S~\ref{sec:conc} is devoted to conclusion.
Throughout the present paper, we adopt the concordance $\Lambda$CDM model with
matter density $\Omega_{\rm m}=0.238$, baryon density $\Omega_{\rm
b}=0.042$, dark energy density $\Omega_{\Lambda}=0.762$ with equation of
state parameter $w=-1$, spectral index $n_s=0.958$, the variance of the
density fluctuation in a sphere of radius 8 $h^{-1}$Mpc $\sigma_8=0.76$,
and Hubble parameter $h=0.732$.  These parameters are consistent with
the WMAP 3-year results~\citep{2007ApJS..170..377S}.  In our ray-tracing
simulation, each realization has an area 
of 25 deg$^2$. The detailed description of our ray-tracing simulations 
is described in
\cite{2009ApJ...701..945S}.

\section{Preliminaries}
\label{sec:basics}

\subsection{Real-Space Correlation Function and Its Covariance Matrix}

In this section we briefly review definitions of the cosmic shear
correlation function and its covariance matrix. 

Since the shear field is a spin-2 field, the field at one particular
point on the sky carries two degrees of freedom. 
We can thus define
different correlation functions from the measured shear field. The most
conventionally used functions are given 
 in terms of the lensing power spectrum as
\citep[e.g.][]{2002A&A...396....1S,2008PhR...462...67M}:
\begin{align}
& \xi_{+}(\theta) = \int^{\infty}_{0}\! \frac{l {\rm d}l}{2\pi}~ 
  P_{\kappa}(l)\;J_{0}(l\theta),\label{cor-plus}\\
& \xi_{-}(\theta) = \int^{\infty}_{0}\! \frac{l {\rm d}l}{2\pi}~ 
  P_{\kappa}(l)\;J_{4}(l\theta),\label{cor-minus}
\end{align}
where $P_\kappa(l)$ is the 
convergence power spectrum \citep[see, e.g.][for the definition]{2001PhR...340..291B,2003astro.ph..5089V},
and $J_0(x)$ and $J_4(x)$
are the zeroth and fourth order Bessel functions, respectively.
Observationally $\xi_+(\theta)$ can be
measured by averaging the product of ellipticity components over
all the galaxy pairs that are separated by the angle $\theta$ (also see
Appendix~\ref{sec:recomp}).

In reality the measured $\xi_+$ and $\xi_-$ are contaminated by
systematic errors. Hence it is in practice useful to decompose 
the measured
$\xi_+$ and $\xi_-$ to the lensing-induced $E$-mode (gradient-mode)
correlation function or equivalently the correlation function of the
convergence field, $\xi_\kappa(\theta)$. Another independent $B$-mode 
correlation function can be used to monitor 
residual systematic errors. 
Although $\xi_\kappa$ and $\xi_+$ contain theoretically equivalent
information in
the absence of systematic errors, which both
have the same expression in
terms of $P_\kappa$ (Eq.~\ref{cor-plus}), the two show a slight
difference when measured from a finite-area survey or simulations as we
will show below in detail.  The difference is ascribed to the fact that
the convergence is the projected mass density field, while the shear
field is a quantity arising from the non-local tidal field that is
affected by the mass distribution outside survey area or 
simulation 
area. 
%
Therefore in this paper we focus on $\xi_+$, rather than
$\xi_\kappa$, to study the covariance matrix
as the shear field is a more direct observable from actual
data.

\begin{figure}
\epsscale{1.0} \plotone{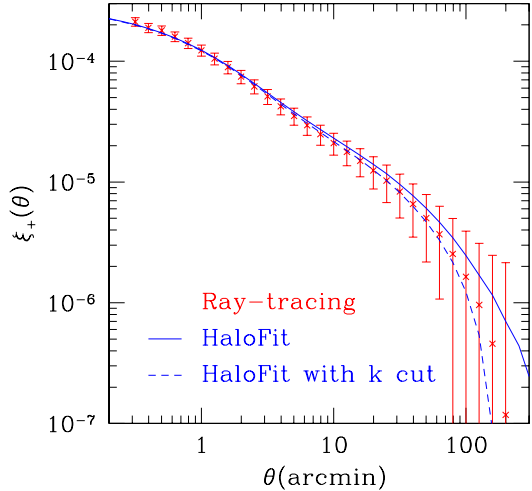} 
\caption{The cosmic shear correlation
function, $\xi_+(\theta)$ (see Eq.~\ref{cor-plus}) for source redshift
$z_s=1.0$, measured from our 1000 ray-tracing simulations. 
The error bars
at each bin are estimated from the scatters ($1\sigma$ scatters) 
in the measured correlation functions
of 1000 simulations. Note that each realization has an area of 25 square
 degrees, therefore the error bars show the sampling variance for the
 area coverage.  
For comparison the solid and dashed curves show the 
analytic predictions computed
using {\em HaloFit}. The difference between the two curves is that the
 dashed curve includes only the contributions of mass density
 fluctuations with length scales within the box size of simulations (see text
 for details). The dashed curve is in better agreement with the
 simulation result over a wide range of separation angles. 
}
\label{tpcf}
\end{figure}

In Fig.~\ref{tpcf} we compare the correlation function $\xi_+(\theta)$,
measured from the shear field in 1000 ray-tracing simulations
\citep[see][for the details of simulations]{2009ApJ...701..945S}, with
the analytical predictions. 
To obtain the analytical predictions we first need to compute the lensing
power spectrum $P_\kappa(l)$, which is given as a projection of the mass
power spectrum weighted with the radial lensing kernel along the line of
sight based on the Limber's approximation:
\begin{equation}
P_\kappa(l)=\int_0^{z_s}\!\!dz~ W_{\rm GL}(z,z_s)P_\delta\!\left(k=
\frac{l}{\chi(z)}; z\right),
\label{eq:pkappa}
\end{equation}
where $W_{\rm GL}(z,z_s)$ is the lensing weight function \citep[see
Eq.~9 in][]{2009ApJ...701..945S} and $P_\delta(k)$ is the mass power
spectrum.  We use the {\em HaloFit} fitting formula
\citep{2003MNRAS.341.1311S} to compute the nonlinear mass power spectrum
$P_\delta(k)$ for the cosmological model we have assumed. In
Fig.~\ref{tpcf} we show the two analytical predictions. One is denoted
by the solid curve showing the prediction computed based on the
conventional method; the two-point correlation function $\xi_+(\theta)$
is computed by inserting the lensing power spectrum, computed from
Eq.~(\ref{eq:pkappa}), into Eq.~(\ref{cor-plus}).  The other is denoted
by the dashed curve, where we properly take into account the fact that
N-body simulations used in ray-tracing simulations do not contain
density perturbations with length scales beyond the simulation box
\citep[see][for details of the simulations]{2009ApJ...701..945S}. For
this purpose we imposed $P_\delta(k; z)=0$ at $k<k_{\rm f}$ and at each
lensing redshift $z$ in computing the lensing power spectrum
(Eq.~\ref{eq:pkappa}), where $k_{\rm f}$ is the fundamental mode of
N-body simulation and given in terms of simulation box size $L$ as
$k_{\rm f}=2\pi/L$. Note that, since our ray-tracing simulations are
done in a light-cone configuration along the line of sight, some
large-length modes are indeed beyond the area of ray-tracing
simulations. 

Fig.~\ref{tpcf} shows that the two analytical predictions are in good
agreement with the simulation results at small separations. However, the
prediction (solid curve) including all the modes beyond simulation box
overestimates the simulation results at large separations,
$\theta\simgt 10'$. On the other hand, the dashed curve, which includes
only a finite box-size effect of the modes, better reproduces the
simulation results at the large separations, up to $\theta\simeq
100'$. Thus these results imply that the two-point correlation function
of a given separation angle is affected by a wide range of Fourier modes
due to the non-local integration relation between the real- and
Fourier-space modes. In particular, even if focusing on the correlation
functions at large separations in the linear regime, we need to properly
take into account an effect of the density perturbations beyond a survey
area, which cannot be observed. 
It is also worth noting
that, if we impose the cutoff on multipole rather than the 3D
wavenumber, e.g. including only the multipole modes at $l>l_{\rm cut}=72$
corresponding to the largest angular mode of our simulated area $25$
sq. degrees, the theoretical prediction underestimates too much the
simulation result at large separation angles.


The results in Fig.~\ref{tpcf} 
can be compared with the result of our previous paper
\citep{2009ApJ...701..945S}, where we compared the simulation and
{\em HaloFit} results for the convergence power spectrum (see Fig.~2 of the
paper). The scale-dependence of the disagreement is qualitatively
different between real- and Fourier-space, reflecting the fact that the
correlation function and the power spectrum is related to each other via
the convolution. From various numerical tests we found that the shear
correlation function remains accurate down to $\theta_{\rm min}\simeq
0.5$ arcmin.
 
Next we move on to the covariance matrix of shear correlation
function. The covariance matrix describes how the correlation functions
of different separation angles are correlated with each other.
 Following the methods developed in \cite{2008A&A...477...43J} and
\cite{2009MNRAS.395.2065T}, the covariance matrix for the correlation
function $\xi_+(\theta)$ is expressed as
\begin{align}
 &{\rm Cov}[\xi_+(\theta),\xi_+(\theta')] =
  \frac{1}{\pi\Omega_{\rm s}}\int_0^{\infty}l {\rm d}l 
  J_0(l\theta)J_0(l\theta')P_{\kappa}(l)^2\nonumber\\ &+
  \frac{1}{4\pi^2\Omega_{\rm s}}\int_0^{\infty}l {\rm d}l\int_0^{\infty}l'
 {\rm d} l'J_0(l\theta)J_0(l'\theta')\bar{T}_{\kappa}(l,l'),
\label{cov of real}
\end{align}
where $\Omega_{\rm s}$ denotes the survey area and $\bar{T}_{\kappa}$
denotes the angle averaged lensing trispectrum.  Note that 
we ignored the shot noise contribution due to random 
intrinsic ellipticities. 
The derivation of Eq.~(\ref{cov of real}) assumes an ideal survey geometry;
in other words Eq.~(\ref{cov of real}) is approximately validated only
for the case, $\theta, \theta'\ll \sqrt{\Omega_{\rm s}}$ and for large
area surveys such as $\Omega_{\rm s}\simgt 1000$ deg$^2$, as discussed
in Appendix~\ref{sec:recomp} in detail.
Therefore it is
expected that for a small survey area the 
covariance matrix 
deviates from the formula above. 
We will call it the
{\em finite area effect} which is examined in Appendix~\ref{sec:recomp}.
See \cite{2008A&A...477...43J} 
for expressions of 
other covariance matrices ${\rm
Cov}[\xi_-(\theta),\xi_-(\theta')]$
and
${\rm Cov}[\xi_+(\theta),\xi_-(\theta')]$, which  
are examined in \S~\ref{sec:F}.

The first term in 
Eq.~(\ref{cov of real}) describes the Gaussian contribution, while the
second term is the non-Gaussian contribution.  There is an important
difference between the covariance of the convergence power spectrum and
that of the real-space correlation function.  Even for a pure Gaussian
field, the first term in Eq.~(\ref{cov of real}) is non-vanishing for
the off-diagonal components with $\theta\ne \theta'$.  The
correlation functions of different separations are always correlated
with each other.  
Also note that the 
covariance
does not depend on the bin width of angles. Thus, when using the
correlation function measurements for constraining cosmological
parameters, it is very important to have an accurate model of the
covariance matrix in order to properly interpret the measurement. 
We will use Eq.~(\ref{cov of
real}) to 
compare the analytical prediction with the covariance 
measured from the simulations.

\subsection{
Constructing a Gaussian Field from the Simulations}
\label{sec:gauss}

The main goal of this paper is to quantify the relative importance of
the non-Gaussian covariance to the Gaussian covariance as a function of
angular scales and source redshifts. 
To address this, we first construct a Gaussian field using the
ray-tracing simulations in order to separate out the Gaussian covariance
contribution. 
We will hereafter 
call the constructed maps ``the simulated Gaussian fields''. 
The reason why 
we use the simulated Gaussian fields 
instead of using the analytical prediction 
(the first term of Eq.~\ref{cov of real}) is as follows. 
Firstly, the ray-tracing simulations do not include large-scale modes beyond the
simulation box size.
Secondly, as we will show below in detail, the
covariance measured from the Gaussian simulations shows a nontrivial
dependence on the survey area that cannot be fully described by the
first term of Eq.~(\ref{cov of real}).

We generated the Gaussian simulations according to the procedures
below. 
Firstly, we Fourier-transformed each convergence field of 1000 ray-tracing
simulations. Secondly, we make a Gaussian field by randomly selecting each
Fourier mode from the Fourier coefficients of 1000 realizations, and then
perform the inverse Fourier transform to obtain the real-space shear field
imposing that the chosen Fourier modes satisfy the
real number condition.
Repeating this procedure, we made 1000 realizations of the Gaussian field. 
The simulated Gaussian fields generated in this way have the same
power spectrum on average 
as that of the original simulations, and
contain
Fourier modes over the same range of angular scales 
as in the original simulations.

\begin{figure}
\epsscale{1.0} \plotone{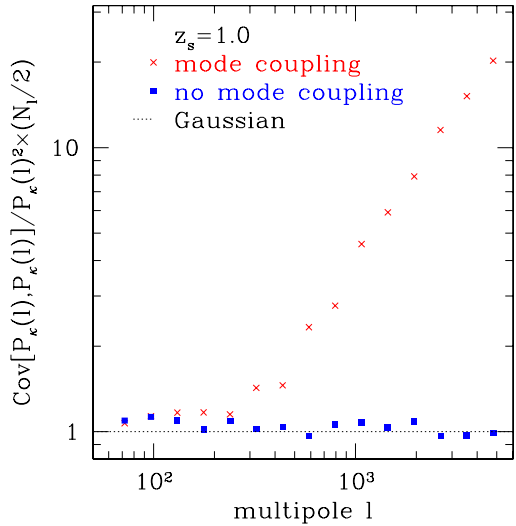} 
\caption{The diagonal
components of the convergence power spectrum covariance, divided by the
expected Gaussian covariance \citep[e.g. see,][]{2009ApJ...701..945S},
 as a function of multipoles. 
 The deviations from unity arise from the non-Gaussian errors.  Note
 that the source redshift is $z_s=1.0$.  The cross symbols are the simulation
 results from 1000 realizations, 
 while the square symbols are the results
 obtained from the Gaussian simulations
we generated from the
  1000 simulations (see \S~\ref{sec:gauss}). The Gaussian simulation
 results are consistent with unity over a range of multipoles.
} 
\label{no_modecoup}
\end{figure}

A justification of the Gaussian fields is given in
Fig.~\ref{no_modecoup}, which shows the diagonal terms of power spectrum
covariance matrix, measured from the original and Gaussian simulated
maps, as a function of multipoles.
The
simulation results are plotted relative to the expectation for a
Gaussian field, where the Gaussian covariance is equal to the squared
power spectrum divided by the number of Fourier modes that are confined
in each multipole bin.
Therefore the deviations from unity arise from the non-Gaussian
error contributions.
The figure explicitly shows that, while the original simulations display
stronger non-Gaussian covariances with increasing multipoles, the
Gaussian simulations 
 are
consistent with the Gaussian expectation over a range of multipoles we
consider. 

\begin{figure}
\epsscale{1.0} \plotone{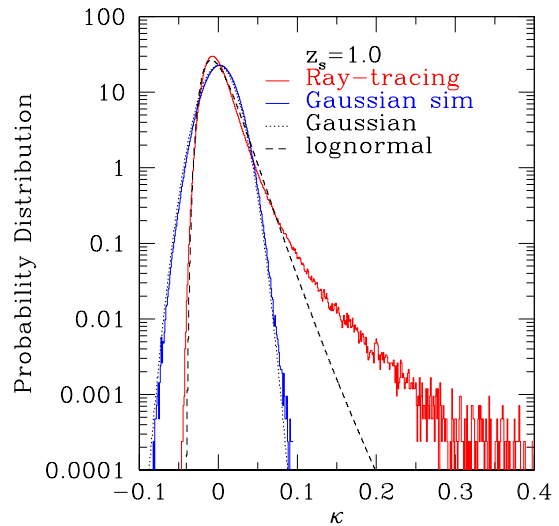} 
\caption{
The probability distributions of
 convergence $\kappa$ computed from the original ray-tracing simulation
 and the Gaussianized simulation, respectively (taken from one
 realization of each simulations). 
The Gaussian simulation result
is consistent with the Gaussian distribution that has zero mean and the
same variance as that measured from the simulation.
} 
\label{kappa_pro}
\end{figure}

Another justification 
is  given 
in Fig.~\ref{kappa_pro}. 
The figure shows the probability distributions of convergence $\kappa$
for one realization of the original ray-tracing simulations (red
histogram) 
and the Gaussianized realization (blue histogram), respectively. The
results are for source redshift $z_s=1.0$. 
The 
original simulation, which includes the non-Gaussian contribution, 
shows a skewed distribution, which strongly deviates from the Gaussian
distribution with the same variance width. 
The distribution is better described 
by a log-normal distribution (dashed curve),
as studied in \cite{2002ApJ...571..638T}, but shows 
a larger positive tail.
On the other hand, the Gaussian simulation is in good agreement with the
Gaussian distribution (dotted curve). 

\section{Results: calibrating the non-Gaussian covariances}
\label{sec:results}
\subsection{Diagonal and Off-Diagonal Components of the Covariance Matrix}
\label{sec:dia-off}

In this section we study the covariance matrix of shear correlation
function using 1000 ray-tracing simulations. 

\begin{figure}
\epsscale{1.0} \plotone{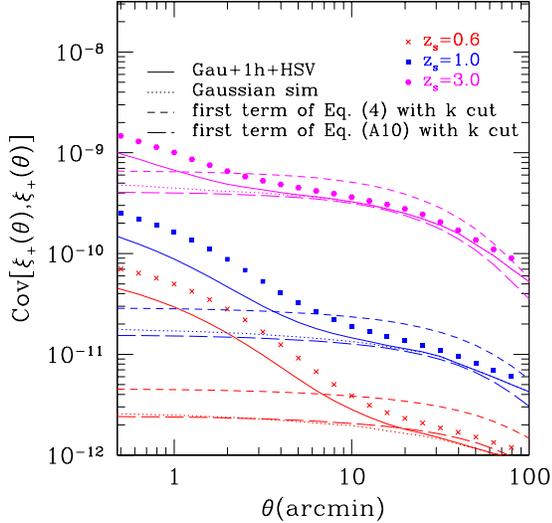} \caption{The diagonal components
of the covariance matrix of shear correlation function for different
source redshifts, $z_s=0.6, 1.0$ and 3.0.  The cross, square and circle
symbols are the simulation results from 1000 realizations for each
source redshifts.  The solid curves show the halo model predictions
including the 1-halo term and the halo sample variance \citep[see][for
the details]{2009ApJ...701..945S}.  The dotted curves are the results
obtained from the Gaussian simulations (see \S~\ref{sec:gauss} and
Fig.~\ref{no_modecoup}).  The short-dashed curves show the analytical
predictions computed from the first term of Eq.~(\ref{cov of real}),
where we included only the Fourier modes contained within our simulation
boxes, i.e. imposed the condition $P_\delta(k)=0$ at $k<k_{\rm f}$ in the
 covariance calculation as in Fig.~\ref{tpcf}. 
On the other hand, the long-dashed curves are the results which are
 computed using an empirical model to account for 
the finite survey area effect (see  Appendix~\ref{sec:recomp} for
 details). This modified Gaussian prediction better matches the Gaussian
 simulation results. 
}  \label{dia_cov}
\end{figure}

The symbols in Fig.~\ref{dia_cov} show the diagonal components of the
covariance matrix measured from 1000 simulations, for different source
redshifts $z_s=0.6, 1.0$ and 3.0, respectively. 
Again notice that each simulation
realization has an area of 25 square degrees, so the plot shows the
covariance expected when measuring the cosmic shear correlation from a
survey with square-shaped survey geometry and 25 square degrees. Here we
ignored the intrinsic ellipticity noise (we will come back to this
later).  First of all, we should stress that, by using the 1000
realizations (25000 square 
degrees in total), we can obtain  well-converged
measurements of the covariance elements over a range of the scales we
consider.
To be more quantitative, 
as shown in Appendix in \cite{2011ApJ...726....7T}, {\em covariances} of
the diagonal covariance elements 
scale approximately as $(2/N_{\rm
r})^{1/2}$, where $N_{\rm r}$ is the number of simulation realizations.
Therefore the covariance elements are measured 
to 4\%-level accuracies.

We first compare the simulation results with the Gaussian error
expectations (short-dashed curve) computed from the first term of
Eq.~(\ref{cov of real}).
Note that we included only the Fourier modes confined within our
simulations in the covariance calculation; 
we imposed $P_\delta(k)=0$ at $k<k_{\rm f}$ as done in
Fig.~\ref{tpcf}. Contrary to the result in Fig.~\ref{tpcf}, the Gaussian
predictions with the $k$-cutoff effect 
appear to overestimate the
simulation results on large separation angles, where the non-Gaussian
errors are insignificant. On the other hands, the dotted curves show the
results obtained from the Gaussian simulations we constructed (see
\S~\ref{sec:gauss}), showing a good agreement with the simulation
results on the large scales.
Hence we conclude that the analytical
prediction given by Eq.~(\ref{cov of real}) is not sufficiently
accurate over the
scales we consider. In fact, as studied in detail in
Appendix~\ref{sec:recomp}, Eq.~(\ref{cov of real}) can be valid only
when the survey area is sufficiently large, such as 1000 square
degrees. 
In Appendix~\ref{sec:recomp} we develop an empirical model to compute
the Gaussian covariance taking into account the finite survey area
effect. 
The long-dashed curve shows the modified analytical prediction for the
Gaussian covariance, which is computed using Eq.~(\ref{new_cov}) as well
as including the
$k$-cutoff for $P_\delta(k)$. The figure
clearly shows that the modified analytical prediction well matches the
Gaussian simulation results.  Thus we need to properly account for both
the finite range of Fourier modes and 
the
finite survey area effect in order to obtain an accurate prediction of
the covariance at large separations.

The simulation results (symbols) start to deviate from the Gaussian
simulation results on small separation angles due to the non-Gaussian
error contribution. The plot shows that the non-Gaussian errors are
more significant on smaller angles and 
for lower source redshifts, because of stronger nonlinear clustering in
structure
formation at lower redshifts and on smaller length scales.  
For comparison, the solid curves show
the halo model prediction developed in our previous paper
\citep{2009ApJ...701..945S}.  To be more precise, the halo model is used
to compute the non-Gaussian covariance for the assumed cosmological
model, including the ``{\em halo sampling variance term}''
\citep{2009ApJ...701..945S} in addition to the 
non-Gaussian term
(1-halo term).
Then the solid curves are the sum of the halo-model-computed
non-Gaussian covariance and the
Gaussian covariance that is computed from the Gaussian simulations. The
halo model predictions qualitatively well reproduce the simulation
results on small separations as well as the source-redshift
dependence. However, the halo model also shows a sizable disagreement
due to the limitation.

\begin{figure*}
\epsscale{1.2}
\plotone{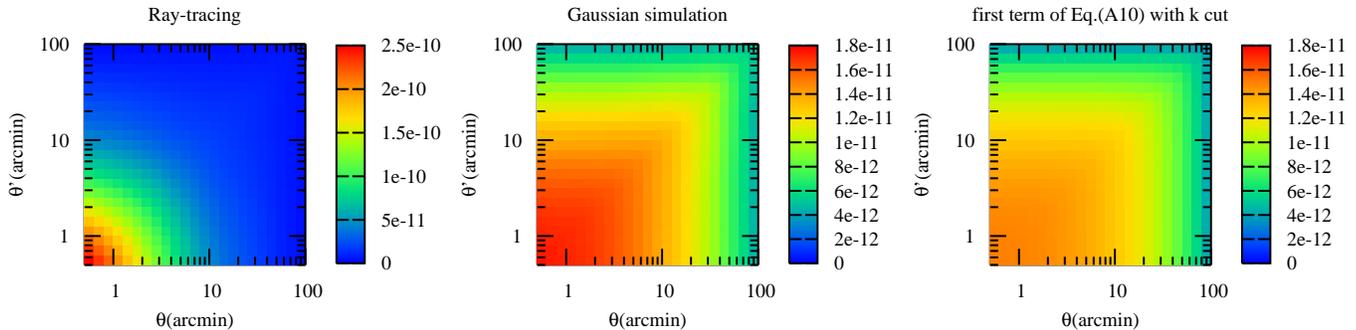}
\caption{
 The covariance matrices ${\rm Cov[\xi_+(\theta),\xi_+(\theta')]}$
 as a function of separation angles $\theta$ and $\theta'$, for the
 ray-tracing simulation realization ({\em left panel}), the Gaussian
 simulation ({\em middle}) and the analytical prediction computed from the
 first term of Eq.~(\ref{cov of real}) ({\em right}), respectively. All the
 results are for source redshift $z_s=1.0$. For the analytical Gaussian
 covariance, we included the finite survey area effect as well as the
 $k$-cutoff in the power spectrum calculation, as in Fig.~\ref{dia_cov}. 
}
\label{2dim_pp}
\end{figure*}

Fig.~\ref{2dim_pp} shows the covariance matrices comparing the results for
the original ray-tracing simulations, the Gaussian simulations and the
Gaussian prediction computed from the first term of Eq.~(\ref{cov of real}), 
where the $k$-cutoff and the finite survey area effect are included.  
Note that all the results are for $z_s=1$. 
The Gaussian prediction (right panel) and the Gaussian simulation
results show a nice agreement for the scale-dependences and 
amplitudes. 
The original simulation results show totally different scale-dependences
from the Gaussian results and display
greater amplitudes at smaller
separation angles (left-lower corner) due to stronger non-Gaussian
contributions. 

\begin{figure*}
\epsscale{1.0} \plotone{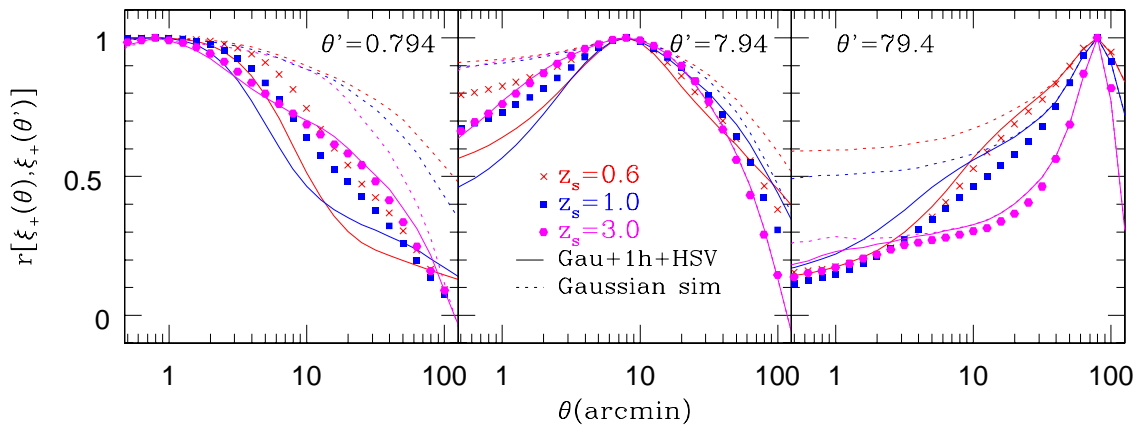} 
\caption{The correlation coefficients $r[\xi_+(\theta),\xi_+(\theta')]$
as a function of separation angles $\theta$, where $\theta'$ is kept
fixed to $\theta'=0.794$, 7.94 and 79.4 (arcmin) in the left, middle and
right panels, respectively.  In each panel we show the results for
different redshifts, $z_s=0.6, 1.0$ and 3.0. The solid curves denote the
halo model predictions, while the dotted curves denote the Gaussian
error predictions.}  \label{off_cov}
\end{figure*}

Next we study the correlation coefficients to quantify strengths of the
off-diagonal covariance components relative to the diagonal components:
\begin{equation}
 r[\xi_+(\theta),\xi_+(\theta')]
\equiv
\frac{{\rm
  Cov}[\xi_+(\theta),\xi_+(\theta')]}{\sqrt{{\rm
  Cov}[\xi_+(\theta),\xi_+(\theta)]{\rm
  Cov}[\xi_+(\theta'),\xi_+(\theta')]}}.
\label{corcoeff}
\end{equation}
The correlation coefficient is defined so that $r=1$ for the diagonal
components when $\theta=\theta'$.
For off-diagonal components when $\theta\neq\theta'$, $r\rightarrow 1$
implies strong correlation between the two correlation functions of
different angular scales, while $r\rightarrow 0$ means no correlation.

Fig.~\ref{off_cov} shows the correlation coefficients
$r[\xi_+(\theta),\xi_+(\theta')]$.
There are 
significant correlations between 
different separation angles.  Comparing the dotted and solid curves
manifests less significant cross-correlations in the non-Gaussian errors
than in the Gaussian errors, implying that the non-Gaussian errors
preferentially contribute to the diagonal components.

\subsection{A Fitting Formula of the Non-Gaussian Covariance Elements}
\label{sec:F}

By using the simulation results shown 
up to the preceding section we derive a
fitting formula to compute the non-Gaussian covariance as a function of
separation angles and source redshifts.  To do this we employ the method
developed in \cite{2007MNRAS.375L...6S}, which is to derive the
calibration function that gives the non-Gaussian covariance contribution
relative to the Gaussian covariance. Our results bring several
improvements over the results in
\cite{2007MNRAS.375L...6S}. Firstly, we use 1000 ray-tracing
realizations, for each source redshift ($z_s=0.6,0.8, 1.0, 1.5, 2.0,
3.0$). Secondly, we carefully estimated the Gaussian covariance
contribution using the Gaussian simulations (see \S~\ref{sec:gauss}),
thereby enabling us to reliably estimate the relative contribution of
the non-Gaussian covariance.

We define $F(\theta,\theta'; z_s)$, the ratio of the non-Gaussian covariance
relative to the Gaussian expectation for the covariance matrix:
\begin{align}
 &F(\theta,\theta'; z_s)
\equiv 
\frac{{\rm
 Cov_{\rm NG}}[\xi_+(\theta),\xi_+(\theta'); z_s]}
{{\rm Cov_{\rm G}}[\xi_+(\theta),\xi_+(\theta'); z_s]}.
\label{calibration}
\end{align}
Here we used the 1000 simulations to compute the Gaussian and
non-Gaussian covariance matrices appearing in the numerator and
denominator of the above equation. 
As shown in Figs.~\ref{dia_cov} and \ref{2dim_pp}
the non-Gaussian contributions are important only on small separation
angles, $\simlt 10$ arcmin, for a $\Lambda$CDM cosmology and for source
redshifts we consider. 

\begin{figure}
\epsscale{1.0}
\plotone{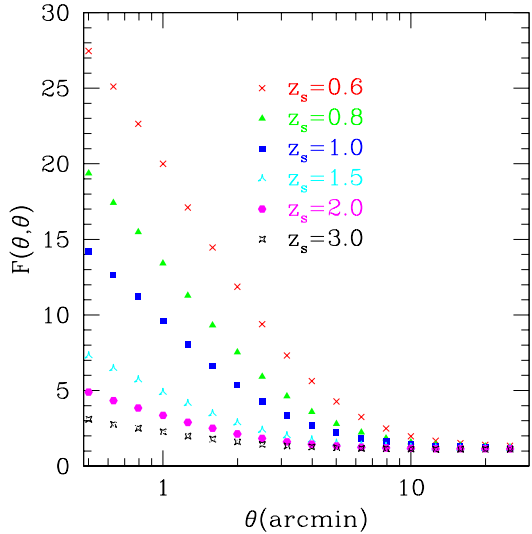}
\caption{The diagonal elements of shear correlation covariance relative
 to the Gaussian covariance contribution, as a function of separation
 angles and source redshifts
(see Eq.~\ref{calibration} for the definition). 
The results are measured from the 
1000 realizations, where the Gaussian contributions are estimated from
 the 1000 Gaussian simulations as in Fig.~\ref{no_modecoup}. }
\label{f_dia}
\end{figure}

Fig.~\ref{f_dia} shows the simulation results for the diagonal
components of $F(\theta,\theta')$ for different source redshifts.  The
fitting formula for $F(\theta,\theta')$ is given in
Appendix~\ref{sec:fit-cal}. 
The figure shows that 
all the curves go to unity, $F(\theta,\theta)=1$, on very large
separations as expected; the non-Gaussian covariances become negligible
on such large separations. This can be contrasted to the results in
\cite{2007MNRAS.375L...6S} (see Fig.~1 in their paper), where the
corresponding curves do not go to unity, even go below unity at
large separations. This is because we carefully computed the Gaussian
covariance contribution (see \S~\ref{sec:gauss}), while
\cite{2007MNRAS.375L...6S} used the analytical prediction (the first
term of Eq.~\ref{cov of real}) to estimate the Gaussian covariance, which
turns out to overestimate the simulation results.
Also note that in \cite{2007MNRAS.375L...6S} all the Fourier modes setting
the lower bound to $k=0$ or $l=0$ are included in the Gaussian
covariance calculation (Eq.~\ref{cov of real}).

The figure also shows that the non-Gaussian errors are significant on smaller
separations and for lower source redshifts, where nonlinear clustering
is more evolving. For a source redshift $z_s\sim 1$, which is a typical
depth of the Subaru-type survey, $F\simgt 10$ on scales smaller than 1
arcminutes, meaning that the non-Gaussian contribution to the diagonal
covariance is greater than the Gaussian contribution by more than a
factor of 10. The non-Gaussian covariance amplitudes
are by accident similar to what is found in
\cite{2007MNRAS.375L...6S}, even though the power spectrum normalization
$\sigma_8$ is quite different: $\sigma_8=0.76$ in our simulations, while 
$\sigma_8=1$ 
 in \cite{2007MNRAS.375L...6S}. Again this is
subscribed to the fact that \cite{2007MNRAS.375L...6S} under-estimated
the non-Gaussian error contribution.
The normalization $\sigma_8=0.76$ we assumed is slightly lower
than the currently most-favored value, $\sigma_8=0.8$ \citep[see
WMAP 7-year result,][]{2011ApJS..192...18K}. 
Therefore the fitting function $F(\theta,\theta')$ we
calibrated may slightly underestimate the non-Gaussian errors 
(by about 20\%). To
correct for this, we can use the halo model prediction to estimate the
difference in the non-Gaussian covariance amplitudes for different
$\sigma_8$ values, and then multiply the correction factor with
$F(\theta,\theta')$. 
 
Let us summarize how to obtain the covariance matrix for given survey
parameters from our fitting formula:
\begin{itemize}
\item[1.] Compute the first term of Eq.~(\ref{new_cov}) 
to estimate the Gaussian
	 covariance matrix, for given survey area and source redshift.
\item[2.] Use the fitting function Eq.~(\ref{fitting}) for
	 $F(\theta,\theta';z_s)$ to obtain the non-Gaussian covariance
	 contribution. 
\item[3.] Multiply the quantities in the steps 1 and 2 to obtain the
	  total covariance matrix. 
\end{itemize}
Exactly speaking, the survey-area dependence of the non-Gaussian 
covariance 
is not as naively expected; ${\rm Cov}\propto 1/\Omega_{\rm s}$
does not hold due to the halo sampling variance contribution
\citep{2009ApJ...701..945S}.
However we have found that the
residual dependence is relatively small, and our fitting formula for the
covariance matrix is approximately valid for a survey area we are most
interested in ($\simgt $100 deg$^2$).  

While we have so far ignored the intrinsic ellipticity contribution to
the covariance, we need to further include the effect.  First of all,
the intrinsic ellipticity noise contributes only to the Gaussian
covariance, and does not affect the non-Gaussian covariance, as long as
the intrinsic ellipticity alignment can be ignored.  Hence we can easily
include the shot noise contribution as follows. 

One method is a
simulation based method. We can generate the Gaussian field including
the shot noise contribution by replacing the power spectrum $P_\kappa(l)$
with 
$P_\kappa(l)+\sigma_\epsilon^2/\bar{n}_{\rm g}$ in generating the
simulation field, where $\sigma_{\epsilon}$ is the rms intrinsic
ellipticity per component and $\bar{n}_g$ is the mean number density of
source galaxies. Here we need to assume $\sigma_\epsilon $ and
$\bar{n}_g$ that we measure from galaxies in a given survey. If we
measure the covariance matrix from the Gaussian simulations generated in
this way, the covariance matrix includes 
 the contribution arising
from the term $\propto (\sigma_\epsilon^2/\bar{n}_g)^2$ as well as 
the mixed term between the
cosmological Gaussian field and the shot noise term arising from the
term $\propto P_\kappa(l)\sigma_\epsilon^2/\bar{n}_g$
\citep{2002A&A...396....1S,2008A&A...477...43J}.

Another way is  following the method in
\cite{2008A&A...477...43J} \citep[also see][]{2002A&A...396....1S},
which developed an 
analytical formula of the Gaussian covariances including the
intrinsic ellipticity noise contribution. 
To be more explicit, the formula is given in terms of the convergence
power spectrum $P_\kappa(l)$, as in the first term of Eq.~(\ref{cov of
real}). Then the intrinsic noise  contribution can be incorporated by
replacing 
$P_{\kappa}(l)^2$
with $
(P_{\kappa}(l)+\sigma_{\epsilon}^2/\bar{n}_g)^2$.
Here we need to assume $\sigma_\epsilon $ and
$\bar{n}_g$ that are measured from
a given survey.  

\begin{figure}
\epsscale{1.0}
\plotone{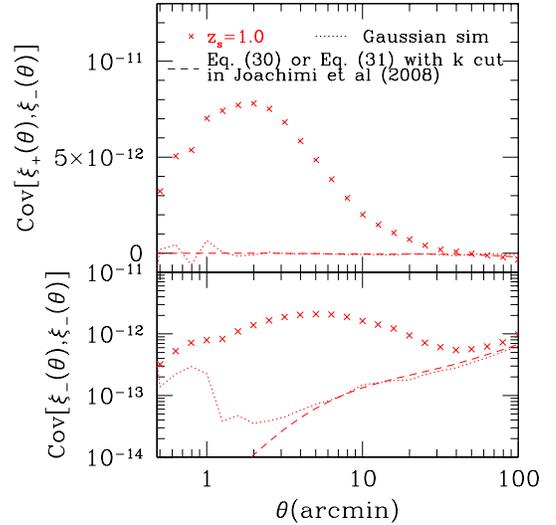}
\caption{The diagonal components of shear correlation covariances
 ${\rm Cov}[\xi_-(\theta),\xi_-(\theta')]$ and ${\rm Cov}[\xi_+(\theta),\xi_-(\theta')]$
as a function of separation angles at source redshift $z_s=1.0$.
The cross symbols are the simulation results measured from the 
1000 realizations. The dotted curves are the results obtained from the
 Gaussian simulations (see \S~\ref{sec:gauss}), while the dashed curves the
 theoretical predictions computed from Eq.~(30) or Eq.~(31) in
 \cite{2008A&A...477...43J}, where we included the $k$-cutoff in the
 lensing power spectrum calculation.
}
\label{dia_cov_other}
\end{figure}

Finally we comment on other contributions to the covariance matrix
we have so far ignored, which are the covariance contributions arising
from another shear correlation function $\xi_-(\theta)$:  
${\rm Cov}[\xi_-(\theta),\xi_-(\theta')]$ and ${\rm
Cov}[\xi_+(\theta),\xi_-(\theta')]$.
Fig.~\ref{dia_cov_other} shows the results. Again the non-Gaussian error
contributions are significant at scales $\simlt 10'$. Compared to
Fig.~\ref{dia_cov}, the covariance matrices 
${\rm Cov}[\xi_-(\theta),\xi_-(\theta')]$ and ${\rm
Cov}[\xi_+(\theta),\xi_-(\theta')]$ have smaller amplitudes than
${\rm Cov}[\xi_+(\theta),\xi_+(\theta')]$ does: for example, 
 the amplitudes of 
${\rm Cov}[\xi_-(\theta),\xi_-(\theta')]$ and ${\rm
Cov}[\xi_+(\theta),\xi_-(\theta')]$ are smaller than that of 
${\rm Cov}[\xi_+(\theta),\xi_+(\theta')]$ by a factor of 100 and 10,
respectively. However, the genuine effects need to be understood in
terms of the signal-to-noise ratios that are roughly estimated as 
 $\xi_+^2/{\rm Cov}[\xi_+,\xi_+]$
or $\xi_+\xi_-/{\rm Cov}[\xi_+,\xi_-]$ at each separation angles.
Since the correlation function $\xi_-$ has smaller amplitudes than 
$\xi_+$
does, by a factor of 10 at separations $\sim 1'$ 
\citep[e.g. see Fig.~2 in][]{2002A&A...396....1S}, therefore the covariances  
${\rm Cov}[\xi_-(\theta),\xi_-(\theta')]$ and ${\rm
Cov}[\xi_+(\theta),\xi_-(\theta')]$ are not negligible.

We tried to derive fitting formulas for the non-Gaussian covariance
contributions to ${\rm Cov}[\xi_-(\theta),\xi_-(\theta')]$ and ${\rm
Cov}[\xi_+(\theta),\xi_-(\theta')]$ in the similar manner as done in
Eq.~(\ref{calibration}). However, due to complex scale-dependences of the
Gaussian covariances as implied in Fig.~\ref{dia_cov_other}, we could
not find useful fitting formulas that are expressed by simple analytical
functions. Therefore, the covariance matrix contributions need to be
directly calibrated from the ray-tracing simulations. The table-format
covariances for ${\rm Cov}[\xi_-(\theta),\xi_-(\theta')]$ and ${\rm
Cov}[\xi_+(\theta),\xi_-(\theta')]$ are available upon 
request. 

\section{Conclusion}
\label{sec:conc}
In this paper, we have developed the theoretical model of the covariance
matrix of cosmic shear two-point correlation function taking into
account the effect of finite survey area and the effects of
non-linear gravitational clustering in large-scale structure.

We found that
the survey-area dependence of the Gaussian covariance,
which scales as $\propto 1/\Omega_{\rm s}$ in the
commonly used formula, 
does not hold for a small area survey due to the effect of finite survey
area. The conventional formula
 is valid only when the survey area is sufficiently wide such as
1000 square degrees.  We examined the residual survey-area dependence
using the method developed by \cite{2002A&A...396....1S}, and obtained
an empirical formula which reproduces our results. 

We examined the non-Gaussian covariance as a function of angular scales and
source redshifts by using two sets of simulation data:
(1) the ray-tracing simulations for the standard $\Lambda$CDM
cosmology and (2) Gaussian fields which have the same power spectrum
on average as that of the original simulations.  
The non-Gaussian errors become more significant on smaller
scales and at lower redshifts.
We compared the simulation results with halo model predictions and found
that the halo model qualitatively well reproduces the non-Gaussian error
over a wide range of separation angles and for redshifts we have considered.
However, the halo model also displays sizable disagreement with the
simulation results. 

Following \cite{2007MNRAS.375L...6S}, we derived the
calibration function to compute the non-Gaussian covariance contribution
relative to the Gaussian covariance (Eq.~\ref{calibration}). 
The Gaussian field data allows us
to cleanly separate out 
the Gaussian covariance contribution accurately,
thereby enabling us to reliably estimate the calibration function.
We found that the calibration factor at arcminute scales can be high as
$\sim$ 20 and $\sim$ 10 for source redshifts of $z_s=0.6$ and $1.0$,
respectively.  
The transition between Gaussian and non-Gaussian covariance occurs
 around 10 and 5 arcminute for $z_s=0.6$ and 1.0, respectively.
Therefore, when one derives constrains on the cosmological parameters
from cosmic shear correlation measurements, it is important to properly
account for the non-Gaussian effects.
The fitting formulae, Eq.~(\ref{new_cov}) and
Eq.~(\ref{fitting}), developed in this paper allow one to compute the
covariance matrix including the non-Gaussian contribution for given
survey parameters (the survey area and the source redshift).

Simulation data (1000 convergence power spectra and cosmic shear
correlation functions for $\xi_{+}(\theta)$ and $\xi_{-}(\theta)$.)
 are available upon request~(contact
\href{mailto:masanori@a.phys.nagoya-u.ac.jp}{masanori@a.phys.nagoya-u.ac.jp}).

\acknowledgments
We thank Naoki Yoshida and Ryuichi Takahashi for useful discussion in
the early stage of this work.  We also thank the anonymous referee for a
very careful reading of our manuscript and very useful and constructive
suggestions, which have helped to improve the manuscript.  M.S. is
supported by the JSPS.  This work is supported in part by World Premier
International Research Center Initiative (WPI Initiative), and
Grant-in-Aid for Scientific Research on Priority Areas No. 467 ``Probing
the Dark Energy through an Extremely Wide and Deep Survey with Subaru
Telescope'' and by the Grant-in-Aid for Nagoya University Global COE
Program, ``Quest for Fundamental Principles in the Universe: from
Particles to the Solar System and the Cosmos'', from the MEXT of Japan.
Numerical computations were in part carried out on the general-purpose
PC farm at Center for Computational Astrophysics, CfCA, of National
Astronomical Observatory of Japan.


\appendix

\section{Finite Area Effect of the Gaussian Covariance}
\label{sec:recomp} 

In this section, we study the validity of the first
term of Eq.~(\ref{cov of real}), which gives the Gaussian error
prediction for the shear correlation function covariance, and will show
the prediction is only valid for a large-area survey covering more than
$1000$ square degrees. For this purpose we will use the method developed
in \cite{2002A&A...396....1S}. 

Let us begin with considering an estimator of the shear
correlation function of a separation angle $\theta$, $\hat{\xi}_+(\theta)$.
For a given galaxy catalog the shear correlation function can be
estimated as
\begin{equation}
 \hat{\xi}_+(\theta)\equiv\frac{1}{N_{\rm
  pair}}\sum_i\sum_j\left(\epsilon_{t(i)}\epsilon_{t(j)}+\epsilon_{\times(i)}\epsilon_{\times(j)}\right)\Delta(|\bm{\theta}_i-\bm{\theta}_j|;\theta),
\label{est-cor}
\end{equation}
where we have used the abbreviated notations such as $\epsilon_{t(i)}$
to denote the ellipticity component of the $i$-th galaxy at the angular
position $\bm{\theta}_i$, 
$N_{\rm pair}$ is the total number of pairs of galaxies that are
separated by 
the separation angle $\theta$ and the index $i$ in summation runs
over all the galaxies in the catalog. 
The tangential and cross components of the ellipticity
$\epsilon=\epsilon_1+{\rm i}\epsilon_2$ at position $\bm{\vartheta}$ are
defined as 
\begin{equation}
 \epsilon_t=-\mathcal{R}e\left(\epsilon\,e^{-2{\rm
			  i}\varphi}\right);\qquad\qquad\epsilon_{\times}=-\mathcal{I}m\left(\epsilon\,e^{-2{\rm
			  i}\varphi}\right),
\end{equation}
where $\mathcal{R}e$ and $\mathcal{I}m$ denote the real- and
imaginary-parts of the quantities and $\varphi$ is the polar angle of
the separation vector between two galaxies, $\bm{\theta}$.  The
component, $\epsilon_t$, is defined as the ellipticity component in
parallel or perpendicular direction relative to the line connecting the
two galaxies. On the other hand $\epsilon_\times$ are measured from the $\pm
45^\circ$ rotated components.
 The function
$\Delta(|\bm{\theta}_i-\bm{\theta}_j|;\theta)$ is a selection function
defined in that $\Delta(|\bm{\theta}_i-\bm{\theta}_j|;\theta)=1$ if
$\theta-\delta\theta/2\le|\bm{\theta}_i-\bm{\theta}_j|\le\theta+\delta\theta/2$,
otherwise $\Delta(|\bm{\theta}_i-\bm{\theta}_j|;\theta)=0$, where
$\delta\theta$ is the bin width of separation angle. The total number of
galaxy pairs is given as $N_{\rm
pair}=\sum_{ij}\Delta(|\bm{\theta}_i-\bm{\theta}_j|;\theta)$. 
The ensemble average of the estimator Eq.~(\ref{est-cor}) is found to indeed
give the shear correlation function: 
\begin{equation}
\langle\hat{\xi}_+(\theta)\rangle=\frac{1}{N_{\rm pair}}\sum_{ij}
\langle 
\epsilon_{t(i)}\epsilon_{t(j)}+\epsilon_{\times(i)}\epsilon_{\times(j)}
\rangle
\Delta(|\bm{\theta}_i-\bm{\theta}_j|;\theta)
=
\frac{1}{N_{\rm pair}}\sum_{ij}
\xi_+(|\bm{\theta}_i-\bm{\theta}_j|)
\Delta(|\bm{\theta}_i-\bm{\theta}_j|;\theta)=\xi_+(\theta). 
\end{equation}

Similarly 
the covariance is defined in terms of the estimator $\hat{\xi}_+(\theta)$ as
\begin{equation}
 {\rm Cov}[\xi_+(\theta),\xi_+(\theta')]=\langle\hat{\xi}_+(\theta)\hat{\xi}_+(\theta')\rangle-\xi_+(\theta)\xi_+(\theta').
\end{equation}
For simplicity, let us consider the diagonal parts of the covariance
matrix, $\theta=\theta'$.
For the Gaussian field, the diagonal parts are computed as
\citep{2002A&A...396....1S}:
\begin{align}
 &{\rm Cov}[\xi_+(\theta),\xi_+(\theta)]=\frac{1}{N_{\rm
 pair}^2}\sum_{ijlm} \langle\left(\epsilon_{t(i)}\epsilon_{t(j)}+
 \epsilon_{\times(i)}\epsilon_{\times(j)}\right)
 \left(\epsilon_{t(l)}\epsilon_{t(m)}+
 \epsilon_{\times(l)}\epsilon_{\times(m)}\right)
 \rangle\Delta(|\bm{\theta}_i-\bm{\theta}_j|;\theta)\Delta(|\bm{\theta}_l-\bm{\theta}_m|;\theta)-\xi_+(\theta)^2\nonumber\\
 &=\frac{1}{N_{\rm
 pair}^2}\sum_{ijlm}[\xi_+(|\bm{\theta}_i-\bm{\theta}_m|)\xi_+(|\bm{\theta}_j-\bm{\theta}_l|)+\cos\left[4\left(\varphi_{im}-\varphi_{jl}\right)\right]\xi_-(|\bm{\theta}_i-\bm{\theta}_m|)\xi_-(|\bm{\theta}_j-\bm{\theta}_l|)]\Delta(|\bm{\theta}_i-\bm{\theta}_j|;\theta)\Delta(|\bm{\theta}_l-\bm{\theta}_m|;\theta)\nonumber\\
 &\equiv\frac{1}{N_{\rm
 pair}^2}\sum_{ijlm}F_{ijlm},
\label{sch.2002}
\end{align}
where $\varphi_{im}$ is the polar angle of the difference vector
$\bm{\theta}_i-\bm{\theta}_m$, and we used the fact that 
 an estimator for $\xi_-(\theta)$ is analogously defined as
\begin{equation}
 \hat{\xi}_-(\theta)\equiv\frac{1}{N_{\rm
  pair}}\sum_i\sum_j\left(\epsilon_{t(i)}\epsilon_{t(j)}
-\epsilon_{\times(i)}\epsilon_{\times(j)}
\right)\Delta(|\bm{\theta}_i-\bm{\theta}_j|;\theta).
\end{equation}
%

Now we use Eq.~(\ref{sch.2002}) to study the effect of finite survey
area. This can be done by comparing the result with the first term of
Eq.~(\ref{cov of real}) because Eq.~(\ref{cov of real}) 
ignores the survey geometry effect. 
To do this we performed the simplified test.
Firstly, we randomly distribute galaxy positions with square shape geometry.
Then, rather than working on the shear field in
simulations, we will compute the summation in Eq.~(\ref{sch.2002}) by
using the tabulated data of $\xi_{+}(\theta)$ and $\xi_{-}(\theta)$.
We estimate $\xi_{+}(\theta)$ and $\xi_{-}(\theta)$ using
Eq.~(\ref{cor-plus}) and Eq.~(\ref{cor-minus}).
In this case, the summation such as $\sum_{ij}$ runs over all the
galaxies in the square-shaped mock simulation, and the separation angle such as
$|\bm{\theta}_i-\bm{\theta}_j|$ can be exactly
 computed from the separation between
the two galaxies chosen.

\begin{figure}
\epsscale{0.4} \plotone{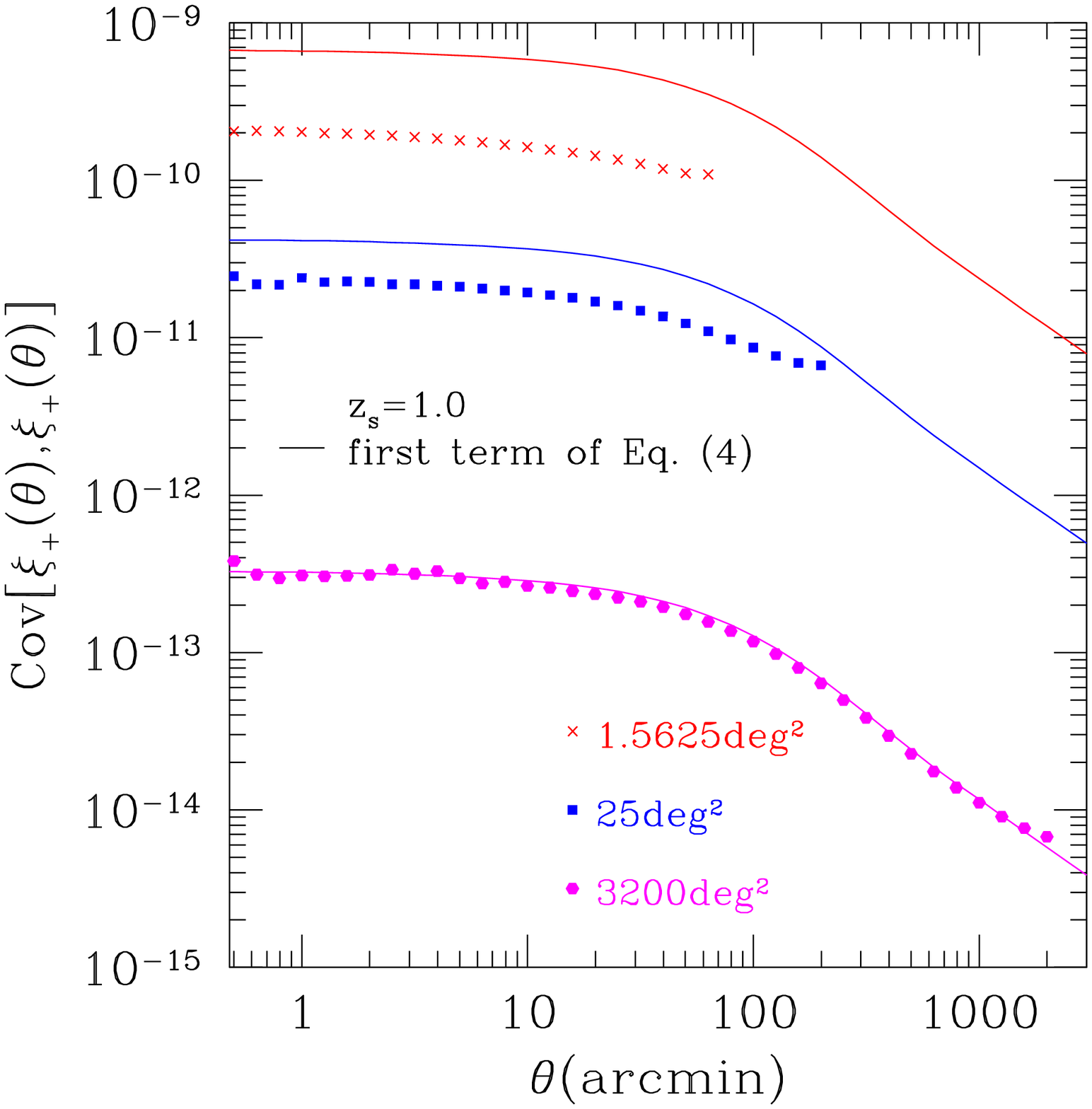} 
\caption{A numerical test for studying the finite area effect on the
Gaussian covariance of shear correlation (see text for the details). 
The symbols show the results
for the diagonal covariance elements, which are computed using
Eq.~(\ref{sch.2002}), for different areas of square-shaped mock simulations:
 $\Omega_{\rm s} = 1.5625$ ${\rm deg}^2$, 25 deg$^2$ and 3200 deg$^2$.  
The source redshift is $z_s=1.0$. 
The solid curves are the prediction computed from the first term of Eq.~(\ref{cov of real}).
The two results agree for a large area such as $\Omega_{\rm s}=3200$
 deg$^2$, but disagree for the smaller areas probably due to the
 finite-area effect. 
}
\label{ffact}
\end{figure}

The symbols in Fig.~\ref{ffact} show the results of Eq.~(\ref{sch.2002})
for the diagonal covariance matrix elements as a function of separation
angles and mock simulation areas. The solid curves are the results computed
from the first term of Eq.~(\ref{cov of real}). In both cases we used
the {\em HaloFit} prediction to compute the lensing power spectrum,
which is then used to compute the shear correlation function as well as
to compute the first term of Eq.~(\ref{cov of real}).
Note that we do not plot the symbols where separation angle becomes
comparable with the scale of the mock simulation area due to avoid the
boundary effect of the square-shaped mock simulation. 

The figure clearly shows that the symbol is in good agreement with the
solid curve for the largest area we consider, $\Omega_{\rm s}=3200$
deg$^2$, however, the two results disagree in amplitudes for the smaller
areas, although the shape looks similar. In addition, comparing the
results for $\Omega_{\rm s}=1.56$ and 25 deg$^2$ manifests that the
disagreement is greater for the smaller survey area. This disagreement
arises due to the effect of the finite survey area.  Hence the
conventionally used expression for the Gaussian covariance, the first
term of Eq.~(\ref{cov of real}), appears to overestimate the covariance
amplitude. The formula is only valid for a sufficiently large area such
as $\Omega_{\rm s}\simgt 1000$ deg$^2$. In other words, due to the
finite survey area effect, the covariance of shear correlation function
does not scale with survey area as ${\rm Cov}\propto 1/\Omega_{\rm s}$,
which is assumed in the conventional formula (Eq.~\ref{cov of real}).

From now on, we will in more detail study 
how the difference between the results of
Eqs.~(\ref{cov of real}) and (\ref{sch.2002}) arises.
Let us consider two survey areas A and B, assuming the area B is
included in the area A (A $\supset$ B).
By using Eq.~(\ref{sch.2002}) the diagonal covariance matrix 
for a survey A can be expressed as 
%
\begin{align}
{\rm Cov}_{\rm A}[\xi_+(\theta),\xi_+(\theta)] &=
\frac{1}{N_{\rm
 pair}^2\mid_{ijlm\in{\rm A}}}\left(\sum_{ijlm\in{\rm
 A-B}}\!\!F_{ijlm}+4\!\!\!\!\sum_{\substack{i\in{\rm B}\\\:{jlm}\in{\rm
 A-B}}}\!\!F_{ijlm}+2\!\!\!\!\sum_{\substack{ij\in{\rm B}\\\quad\!{lm}\in{\rm A-B}}}\!\!F_{ijlm}\right.\nonumber\\
&\left.+\:{2}\!\!\!\!\sum_{\substack{\!\!{il}\in{\rm B}\\\quad\!\!\!\!{jm}\in{\rm A-B}}}\!\!F_{ijlm}
 +2\!\!\!\!\sum_{\substack{\!\!\!\!{im}\in{\rm B}\\\quad{jl}\in{\rm
 A-B}}}\!\!F_{ijlm}+4\!\!\!\!\sum_{\substack{{jlm}\in{\rm B}\quad\\\quad\:{i}\in{\rm A-B}}}\!\!F_{ijlm}+
 \!\!\!\!\sum_{ijlm\in{\rm B}\quad}\!\!F_{ijlm}\right),
\label{covA_B}
\end{align}
where the notation $ij\in{\rm B}$ is introduced to mean that 
the indices $i$ and $j$ are included in B, and the notation 
 $lm\in{\rm A-B}$ means that the indices $l$ and $m$ are
included in the region that is in A, but not in B.
We will consider the case that survey area A is sufficiently large in the 
following discussion.
We define the value arising from finite area effect as
\begin{equation}
{\rm FAE}\equiv\frac{\rm A}{\rm B}{\rm Cov}_{\rm A}[\xi_+(\theta),\xi_+(\theta)]-{\rm Cov}_{\rm B}[\xi_+(\theta),\xi_+(\theta)].
\label{eq:FAE}
\end{equation}
If B is close to A, there is no finite area effect 
and FAE should be zero
because of survey area as ${\rm Cov}\propto{1/\Omega_{\rm s}}$. 
If B is close to A, the final term of Eq.~(\ref{covA_B}) has a dominant
contribution because the number of galaxy
pairs is greater than other terms. 
In this case Eq.~(\ref{eq:FAE}) becomes
\begin{align}
{\rm FAE}&\equiv\frac{\rm A}{\rm B}{\rm Cov}_{\rm A}[\xi_+(\theta),\xi_+(\theta)]-{\rm Cov}_{\rm
 B}[\xi_+(\theta),\xi_+(\theta)]\simeq\frac{\rm A}{\rm B}\frac{1}{N_{\rm
 pair}^2\mid_{ijlm\in{\rm A}}}\sum_{ijlm\in{\rm B}\quad}\!\!F_{ijlm}-{\rm
			 Cov}_{\rm B}[\xi_+(\theta),\xi_+(\theta)]\nonumber\\
&\simeq\frac{1}{N_{\rm
 pair}^2\mid_{ijlm\in{\rm B}}}\sum_{ijlm\in{\rm B}\quad}\!\!F_{ijlm}-{\rm
			 Cov}_{\rm B}[\xi_+(\theta),\xi_+(\theta)]=0.
\end{align}
%
Thus Eq.~(\ref{eq:FAE}) has an asymptotic limit of 
${\rm Cov}_{\rm A}={\rm Cov}_{\rm B}$ when ${\rm A}={\rm B}$.  
However, the other terms in Eq.~(\ref{covA_B}) are not
 negligible if ${\rm A}\ne {\rm B}$, and the covariances for a general
 case do not scale with survey area as 
${\rm Cov}\propto 1/\Omega_{\rm s}$, due to a finite survey area effect. 

\begin{figure}
\epsscale{0.4}
\plotone{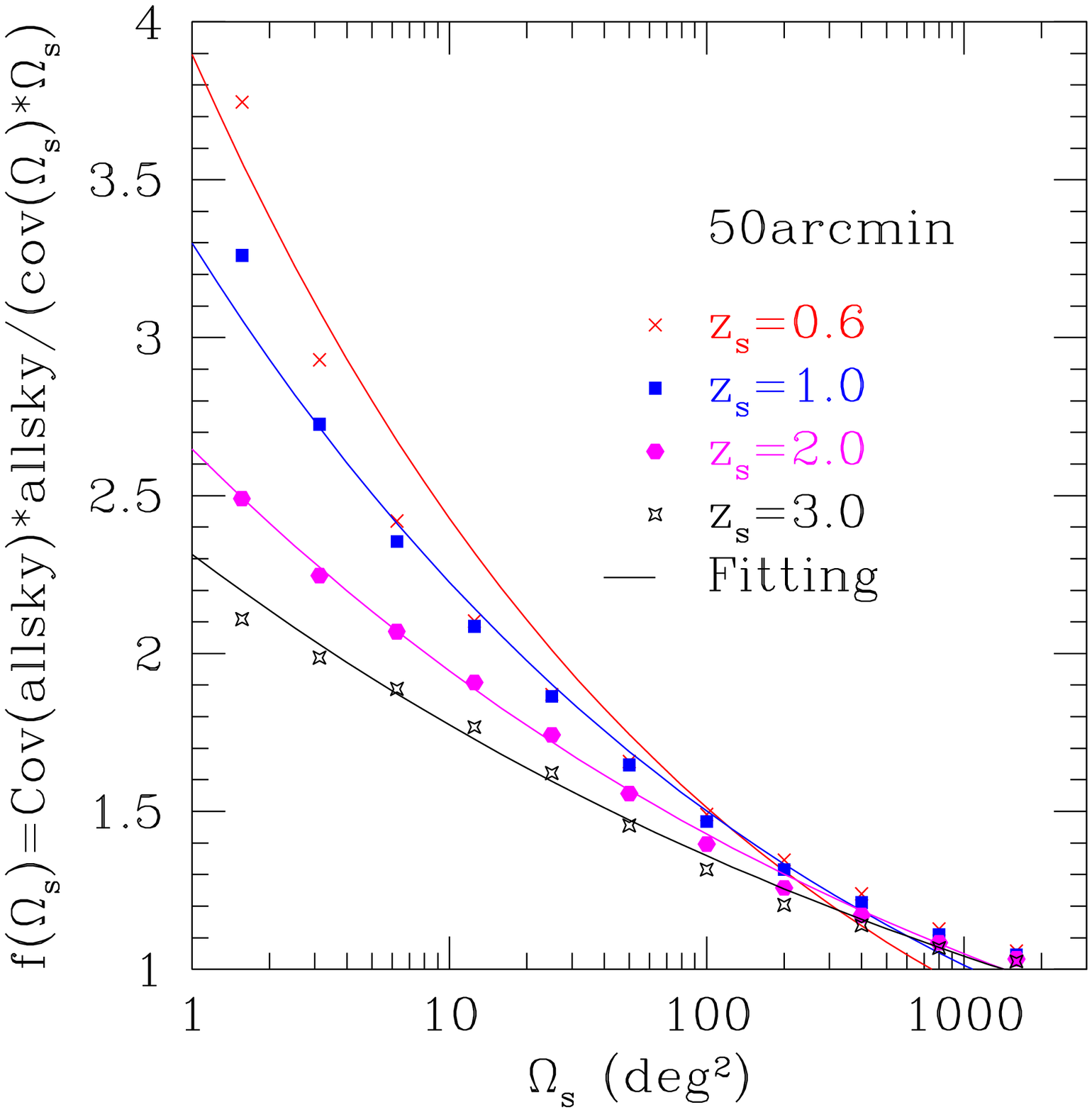}
\caption{A fitting formula for the survey area dependence of the Gaussian
 covariance, $f(\Omega_{\rm s})$ in Eq.~(\ref{new_cov}).
The symbols are the mock simulation results for different source redshifts as
 in Fig.~\ref{ffact}, and the solid curves are the fitting formula given
 by Eq.~(\ref{fitting_v1}). We did the fitting by matching to the
 mock simulation results for the diagonal Gaussian matrix at the scale of
 50 arcmin. 
}
\label{scale}
\end{figure}

Therefore we will estimate a fitting formula to account for the survey
area dependence by assuming the form 
\begin{align}
 {\rm Cov}[\xi_+(\theta),\xi_+(\theta')] =
  \frac{1}{\pi\Omega_{\rm s}f(\Omega_{\rm s})}\int_0^{\infty}l {\rm d}l 
  J_0(l\theta)J_0(l\theta')P_{\kappa}(l)^2+
  \frac{1}{4\pi^2\Omega_{\rm s}f(\Omega_{\rm s})}\int_0^{\infty}l {\rm d}l\int_0^{\infty}l'
 {\rm d} l'J_0(l\theta)J_0(l'\theta')\bar{T}_{\kappa}(l,l'),
\label{new_cov}
\end{align}
where $f(\Omega_{\rm s})$ denotes the new survey area dependence.
We parametrize $f(\Omega_{\rm s})$ as
\begin{equation}
 f(\Omega_{\rm s})=\frac{\alpha(z_s)}{\Omega_{\rm s}^{\beta(z_s)}}.\label{fitting_v1}
\end{equation}
We estimate the 2 parameters  using the mock simulations as in
Fig.~\ref{ffact}, for 
different source redshifts $z_s=0.6,0.8,1.0,1.5,2.0$ and $3.0$.
The mock simulation results are well fitted by the parameters
\begin{align}
 \alpha(z_s)=\alpha_1 z_s^{\alpha_2}\nonumber\\
 \beta(z_s)=\beta_1 z_s^{\beta_2}.
\end{align}
The best-fitting parameters are found to be
$(\alpha_1,\alpha_2)=(3.2952,-0.316369)$ and 
 $(\beta_1,\beta_2)=(0.170708,-0.349913)$, 
respectively. As shown in Fig.~\ref{scale}
we made the fitting over survey areas of
$1.5625\leq\Omega_{\rm s}\leq 1600$ deg$^2$ using the diagonal
covariance at 50 arcmin. However we checked that the above fitting
formula fairly well reproduce the results for different separations and
for the off-diagonal covariance elements.
A caution on the use of the fitting formula is the output value should
be replaced to unity if the value is below unity. 

\section{Fitting Formula for Calibration Matrix}
\label{sec:fit-cal}

\begin{figure}
\epsscale{0.4} \plotone{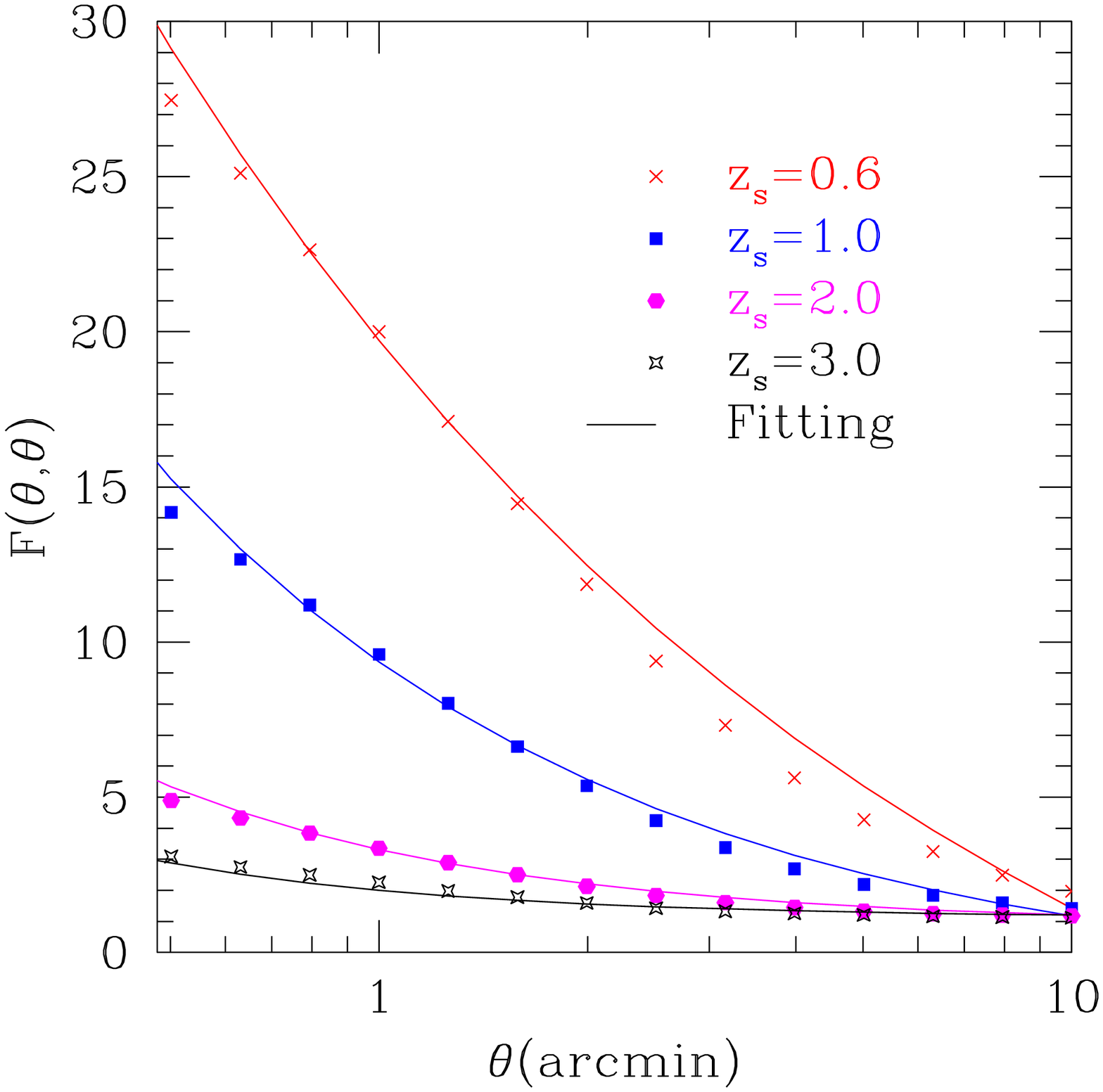} 
\caption{The diagonal components
of the calibration function to compute the non-Gaussian covariance,
$F(\theta,\theta;z_s)$ (see Eq.~\ref{calibration} for the definition).
The symbols are the same as in Fig.~\ref{f_dia}, and the solid curves
 are the fitting formula given by 
Eq.~(\ref{fitting}).}
\label{fit_dia}
\end{figure}

\begin{figure*}
\epsscale{1.0}
\plotone{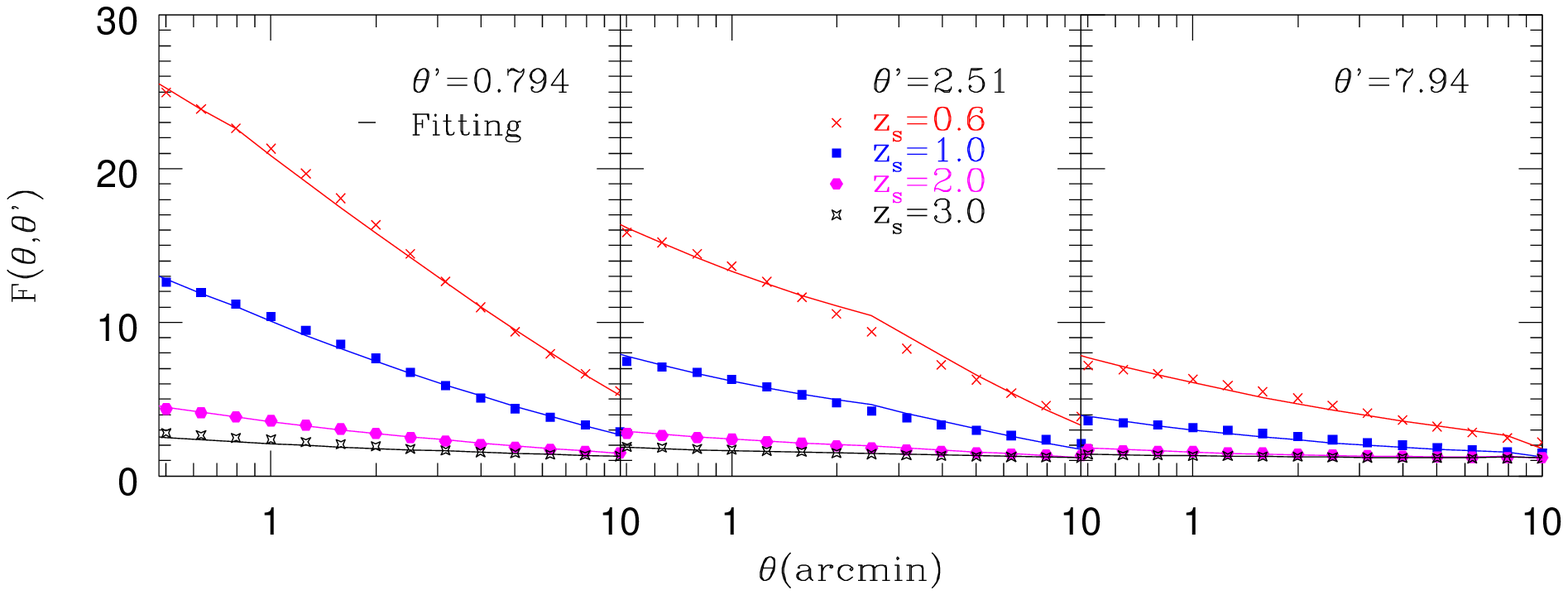}
\caption{As in the previous figure, but for the off-diagonal
 components. Shown is the function $F(\theta,\theta')$ 
as a function of $\theta$, but 
$\theta'$ is kept fixed to $\theta'=0.794$,
 2.51 and 7.94 (arcmin) in the left, middle and right panels,
 respectively.
}
\label{fit_off}
\end{figure*}

In this section we estimate the calibration function $F(\theta,\theta;
z_s)$ we studied in \S~\ref{sec:F}, which can be used to estimate the
non-Gaussian covariance in combination with the fitting formula
(\ref{new_cov}) for the Gaussian covariance.
We parametrize $F(\theta,\theta'; z_s)$ as
\begin{equation}
 F(\theta,\theta'; z_s)=\left(a(z_s)+\frac{b(z_s)}{(\theta\theta')^{c(z_s)}}\right)\cdot
  d(z_s)^{|\theta-\theta'|}.
\label{fitting}
\end{equation}
The 4 parameters are estimated from the ray-tracing simulations for 6
 different source redshifts $z_s=0.6,0.8,1.0,1.5,2.0$ and $3.0$.
As demonstrated in Figs.~\ref{fit_dia} and \ref{fit_off},
 the simulation results are well
fitted by the following best-fitting parameters: 
 \begin{align}
&a(z_s)=-z_s^{a_1}\exp(a_2 z_s)+a_3\nonumber\\
&b(z_s)=b_1 z_s^{b_2}+b_3\nonumber\\
&c(z_s)=c_1 z_s^{c_2}+c_3\nonumber\\
&d(z_s)=d_1 z_s^{d_2}+d_3,
\end{align}
where 
$(a_1,a_2,a_3)=(-3.7683,0.9752,1.4048)$, 
 $(b_1,b_2,b_3)=(10.7926,-2.0284,-0.2266)$,
 $(c_1,c_2,c_3)=(-0.3664,-0.5733,0.6863)$, and
 $(d_1,d_2,d_3)=(0.2450,0.1218,0.7076)$, 
respectively. 
We made this fitting over 
angular scales $0.5<\{\theta,\theta'\}\leq$ 10 arcmin.
Figs.~\ref{fit_dia} and \ref{fit_off} clearly show that the fitting
formula above well reproduce the simulation results, and the accuracy of
the fitting formula is within about 25\%. 
However it should be noted that the fitting formula is only applied to source
redshift ranges $0.6\le z_s\le 3.0$ and angular
 scales below 10 arcmin.
This is sufficient because the non-Gaussian covariance is 
 important only below 10 arcmin.
   
\bibliography{ms}
\clearpage

\end{document}